\documentclass[amsmath,amssymb,aps,reprint]{revtex4-2}

\usepackage{algorithm}
\usepackage[noend]{algpseudocode}
\usepackage{mathtools}
\usepackage{placeins}
\usepackage{subfiles}
\usepackage[hidelinks]{hyperref}
\usepackage{orcidlink}
\usepackage{aligned-overset}
\usepackage{datetime}

\newcommand\numberthis{\addtocounter{equation}{1}\tag{\theequation}}

\newcommand\algorithmicbreak{\textbf{break}}
\newcommand\algortihmiccontinue{\textbf{continue}}

\makeatletter
\def\algbackskip{\hskip-\ALG@thistlm}
\makeatother

\begin{document}

\title{Clock offset recovery with sublinear complexity enables synchronization on low-level hardware for quantum key distribution}

\author{Jan~Krause\,\orcidlink{0000-0002-3428-7025}}
\email[]{jan.krause@hhi.fraunhofer.de}
\author{Nino~Walenta\,\orcidlink{0000-0001-7243-0454}}
\author{Jonas~Hilt}
\author{Ronald~Freund\,\orcidlink{0000-0001-9427-3437}}
\affiliation{Fraunhofer Institute for Telecommunications, Heinrich-Hertz-Institut, HHI, 10587 Berlin, Germany}

\newdate{date}{05}{04}{2024}
\date{\displaydate{date}}

\begin{abstract}
We introduce iQSync, a clock offset recovery method designed for implementation on low-level hardware, such as FPGAs or microcontrollers, for quantum key distribution (QKD).
iQSync requires minimal memory, only a simple instruction set (e.g. no floating-point operations), and can be evaluated with sublinear time complexity, typically involving no more than a few thousand iterations of a simple loop.
Furthermore, iQSync allows for a precise clock offset recovery within few seconds, even for large offsets, and is well suited for scenarios with high channel loss and low signal-to-noise ratio, irrespective of the prepare-and-measure QKD protocol used.
We implemented the method on our QKD platform, demonstrating its performance and conformity with analytically derived success probabilities for channel attenuations exceeding $70\,\mathrm{dB}$.
\end{abstract}

\maketitle
\section{\label{sec:introduction}Introduction}

\begin{figure}[b]
    \includegraphics[width=\linewidth]{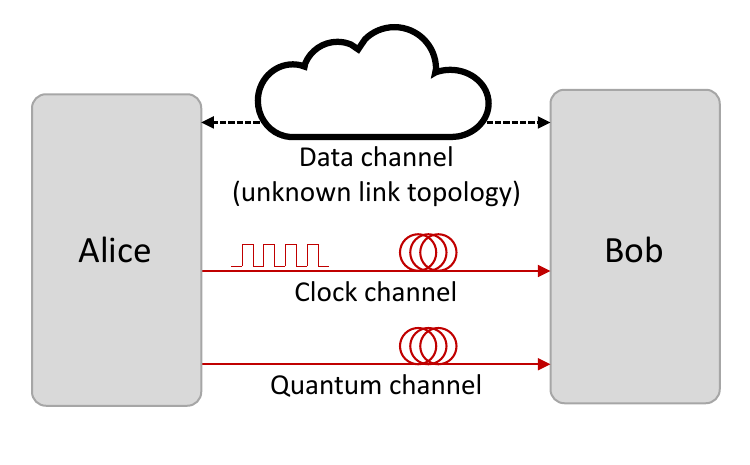}
    \caption{\label{fig:setup}
    A typical QKD setup.
    Qubits, e.g. encoded in coherent quantum states with low mean photon number $\mu \lesssim 1$,
    are transmitted from Alice to Bob using a \textit{quantum channel}.
    Due to channel losses, Bob usually receives states with $\mu \ll 1$.
    A \textit{bi-directional data channel} with high (e.g. a few hundred ms)
    unknown and undeterministic (but bounded) latencies is used for
    classical communication between Alice and Bob, and for QKD post-processing.
    An optional \textit{clock channel} is often used to establish a common clock
    phase.
    }
\end{figure}

Quantum key distribution (QKD) allows for the generation of symmetric cryptographic keys between two remote parties with information-theoretical protocol security
\cite{bennettQuantumCryptographyPublic1984,portmannSecurityQuantumCryptography2022}.
In contrast to conventional optical communication, QKD is based on the transmission and detection of single photons, which complicates the required precise clock synchronization between sender (Alice) and receiver (Bob).
Bob typcially only receives a small fraction of the symbols transmitted by Alice, usually in the range of $1/10$ to $1/10^7$
\cite{boaronSecureQuantumKey2018}.
Additionally, the quantum bit error rate (QBER) can reach up to approximately $11\,\%$
\cite{gottesmanSecurityQuantumKey2004},
substantially exceeding typical bit error rates of $10^{-9}$ or less encountered in classical fiber-optical communication systems
\cite{agrawalFiberopticCommunicationSystems2010}.
These conditions prevent the usage of established synchronization methods from
classical telecommunication and therefore necessitate the usage of different approaches.

Conceptually, the task of clock synchronization between Alice and Bob can
be divided into \textit{clock phase-locking}, i.e. ensuring that both clocks operate at the exact same frequency, and \textit{clock-offset recovery},
needed for Bob to assign the correct qubit index to each detected photon.
Ideally, a synchronization method also provides
the ability to identify large clock offsets,
reliable operation,
high precision,
fast completion within seconds even for high transmission losses,
algorithmic simplicity,
low resource usage,
low cost of the required components,
and no dependence on third-party infrastructure like GPS satellites.

Clock phase-locking can be realized through transmission of Alice's clock beat via a dedicated clock channel
\cite{
  bienfangQuantumKeyDistribution2004,
  tanakaUltraFastQuantum2008,
  liuDecoystateQuantumKey2010,
  walentaFastVersatileQuantum2014,
  korzhProvablySecurePractical2015,
  dynesUltrahighBandwidthQuantum2016,
  islamProvablySecureHighrate2017,
  zhangTimingSynchronisationHigh2021,
  berraSynchronizationQuantumCommunications2023},
by using (GPS-disciplined) atomic clocks
\cite{
  jenneweinQuantumCryptographyEntangled2000,
  marcikicFreespaceQuantumKey2006,
  ursinEntanglementbasedQuantumCommunication2007,
  ervenEntangledQuantumKey2008},
or by exploiting the timing characteristics of single-photon detections, which was demonstrated for entanglement-based systems
\cite{
  hoClockSynchronizationRemote2009,
  fitzkeScalableNetworkSimultaneous2022,
  spiessClockSynchronizationCorrelated2023},
as well as for prepare-and-measure systems
\cite{
  takenakaSatellitetogroundQuantumlimitedCommunication2017,
  calderaroFastSimpleQubitbased2020,
  cochranQubitBasedClockSynchronization2021,
  wangSynchronizationUsingQuantum2021,
  agnesiTimebinQuantumKey2022,
  spiessClockSynchronizationPulsed2023,
  zahidySinglePhotonBasedClockAnalysis2023,
  lohrmannClassicalClockSynchronization2023}.

Clock offset recovery is mostly performed via single-photon detections in the quantum channel with approaches existing for entanglement-based systems
\cite{
  ursinEntanglementbasedQuantumCommunication2007,
  ervenEntangledQuantumKey2008,
  hoClockSynchronizationRemote2009,
  scheidlFeasibility300Km2009,
  eckerStrategiesAchievingHigh2021,
  fitzkeScalableNetworkSimultaneous2022,
  spiessClockSynchronizationCorrelated2023,
  laflerQuantumTimeTransfer2023},
and for prepare-and-measure systems
\cite{
  takenakaSatellitetogroundQuantumlimitedCommunication2017,
  calderaroFastSimpleQubitbased2020,
  cochranQubitBasedClockSynchronization2021,
  wangSynchronizationUsingQuantum2021,
  agnesiTimebinQuantumKey2022,
  spiessClockSynchronizationPulsed2023,
  wangRobustFrameSynchronization2023,
  walentaQuantumCommunicationSynchronization2023}.
While methods employing the classical channel exist
\cite{
  zhangTimingSynchronisationHigh2021,
  berraSynchronizationQuantumCommunications2023},
they introduce additional routing constraints.

One of the simplest approaches for clock offset recovery via single-photon detections in the quantum channel employs a fast Fourier transform (FFT)
\cite{cooleyAlgorithmMachineCalculation1965}
with time complexity (TC) in $\mathcal{O}(n \log_2 n)$, where $n$ corresponds to the maximum recoverable offset.
For typical offsets of up to multiple $100 \, \mathrm{ms}$ and sub-nanosecond symbol durations
\cite{boaronSecureQuantumKey2018},
this imposes high demands on CPU and RAM.
Furthermore, calculation of the FFT requires floating-point operations and substantial resources on platforms like FPGAs.

\begin{figure}
  \includegraphics[width=\linewidth]{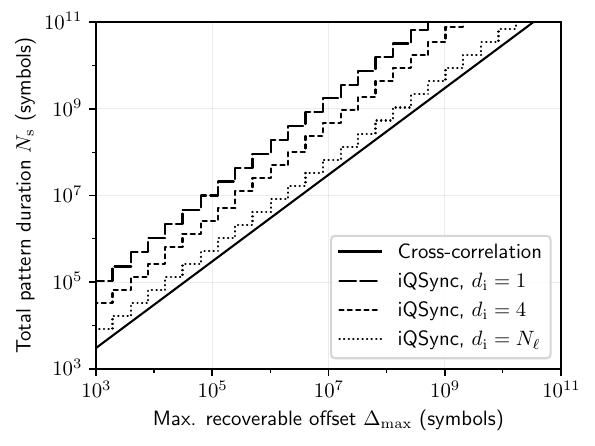}
  \caption{\label{fig:pattern_duration_over_max_offset}
  Relation between the total synchronization pattern duration $N_\mathrm{s}$ and the max. recoverable offset $\Delta_\mathrm{max}$ for different degrees of interleaving $d_\mathrm{i}$.
  Cross-correlation refers to approaches utilizing a single FFT, or the dual-step FFT approach as presented in \cite{calderaroFastSimpleQubitbased2020}.
  }
\end{figure}

An approach with reduced TC of $\mathcal{O}(n \log_2 \log_2 n)$ was presented in 2020 by Calderaro et al.
\cite{calderaroFastSimpleQubitbased2020}.
It uses a random pattern that is constructed such that two FFTs on reduced datasets can be calculated.
However, it still requires FFTs on large datasets and floating-point operations, hindering its usage on low-level hardware.

Another approach, proposed by Braithwaite, uses a dichotomic search algorithm to consecutively reconstruct the bits of the binary representation of the clock offset, starting from the least significant bit
\cite{braithwaiteQuantumChannelSynchronization2020}.
It only requires simple computational instructions, i.e. no floating-point operations, at the expense of a pattern duration that scales super-linearly with the maximum recoverable offset as $\mathcal{O}(n \log_2 n)$.
Furthermore, it requires the exchange of multiple confirmation messages between Alice and Bob.

Here, we propose iQSync, a synchronization method, which is also based on a consecutive bit-wise recovery of the clock offset, but uses random interleaving of the patterns required for the recovery of each bit.
This approach allows for keeping a linear scaling $\mathcal{O}(n)$ of the overall synchronization pattern length, while offset recovery at Bob is reduced to sublinear TC.

The free choice of the degree of interleaving $d_\mathrm{i}$ allows for optimization of resources in any given hardware as well as finetuning of the trade-off between failure probability and required pattern duration.

Furthermore, not more than one single message needs to be transmitted from Alice to Bob at the beginning of the synchronization.
The sychronization pattern can be generated and evaluated without complex calculations, using only few additions and bit-shifts within a loop.
Also, storage of pre-computed patterns is not necessary for our method.
Therefore, our method is ideally suited for cost-efficient implementation on low-level hardware.

Pattern generation and the dichotomic search offset recovery algorithm for iQSync are described in Section~\ref{sec:method} and the success probability is analytically derived in Section~\ref{sec:success_probability}.
The TC of iQSync is analyzed in Section~\ref{sec:computational_complexity}.
An experimental validation of these results is presented in Section~\ref{sec:experiment}.
Finally, Section~\ref{sec:conclusion} concludes this work.
\section{\label{sec:method}Methods}

\begin{table*}
    \caption{\label{tab:symbols-iqsync-level-2-p-1}
    Example symbol pattern for iQSync, using a maximum level $\ell_\mathrm{max} = 2$, and the degree of interleaving $d_\mathrm{i}=1$ (no interleaving).
    Each level consists of $N_\mathrm{s,\ell} = 2^{\ell_\mathrm{max} + 1}$ symbols and is used to reconstruct one bit of the binary integer representation of the clock offset.
    Level 0 is only needed in case a symbol is encoded in more than one timebin, as is the case for our implementation.
    An algorithmic description of the pattern generation is given in Algorithm~\ref{alg:pattern-gen-iqsync}.
    }
    \begin{ruledtabular}
        \begin{tabular}{l|cccccccc|cccccccc|cccccccc}
            Symbol index $k$ & 0 & 1 & 2 & 3 & 4 & 5 & 6 & 7 & 8 & 9 & 10 & 11 & 12 & 13 & 14 & 15 & 16 & 17 & 18 & 19 & 20 & 21 & 22 & 23 \\
            \colrule
            Level $\ell$          & 0 & 0 & 0 & 0 & 0 & 0 & 0 & 0 & 1 & 1 & 1 & 1 & 1 & 1 & 1 & 1 & 2 & 2 & 2 & 2 & 2 & 2 & 2 & 2 \\
            Symbol $s$            & 0 & 0 & 0 & 0 & 0 & 0 & 0 & 0 & 0 & 1 & 0 & 1 & 0 & 1 & 0 & 1 & 0 & 0 & 1 & 1 & 0 & 0 & 1 & 1 \\
        \end{tabular}
    \end{ruledtabular}
  \end{table*}
  \begin{table*}
    \caption{\label{tab:symbols-iqsync-level-3-p-2}
    Example symbol pattern for iQSync, using a maximum level $\ell_\mathrm{max} = 3$, and the degree of interleaving $d_\mathrm{i} = 2$.
    Each group (separated by the vertical line) is constructed from $d_\mathrm{i} = 2$ interleaved levels by randomly choosing the level from which to take the symbol value for each symbol.
    An algorithmic description of the pattern generation is given in Algorithm~\ref{alg:pattern-gen-iqsync}.
    Intermediate steps for offset recovery are given in the lower half and described in Sec.~\ref{sec:pattern-analysis}.
    }
    \begin{ruledtabular}
        \begin{tabular}{l|cccccccccccccccc|cccccccccccccccc}
            Symbol index $k$                              & 0 & 1 & 2 & 3 & 4 & 5 & 6 & 7 & 8 & 9 & 10 & 11 & 12 & 13 & 14 & 15 & 16 & 17 & 18 & 19 & 20 & 21 & 22 & 23 & 24 & 25 & 26 & 27 & 28 & 29 & 30 & 31 \\
            \colrule
            Group index $g$                               & \multicolumn{16}{c|}{0} & \multicolumn{16}{c}{1} \\
            \colrule
            Symbol for $\ell = 0$                   & 0 & 0 & 0 & 0 & 0 & 0 & 0 & 0 & 0 & 0 & 0 & 0 & 0 & 0 & 0 & 0 &   &   &   &   &   &   &   &   &   &   &   &   &   &   &   &   \\
            Symbol for $\ell = 1$                   & 0 & 1 & 0 & 1 & 0 & 1 & 0 & 1 & 0 & 1 & 0 & 1 & 0 & 1 & 0 & 1 &   &   &   &   &   &   &   &   &   &   &   &   &   &   &   &   \\[1.5ex]
            Symbol for $\ell = 2$                   &   &   &   &   &   &   &   &   &   &   &   &   &   &   &   &   & 0 & 0 & 1 & 1 & 0 & 0 & 1 & 1 & 0 & 0 & 1 & 1 & 0 & 0 & 1 & 1 \\
            Symbol for $\ell = 3$                   &   &   &   &   &   &   &   &   &   &   &   &   &   &   &   &   & 0 & 0 & 0 & 0 & 1 & 1 & 1 & 1 & 0 & 0 & 0 & 0 & 1 & 1 & 1 & 1 \\[1.5ex]
            Randomly chosen $\ell$                  & 0 & 0 & 1 & 0 & 1 & 1 & 1 & 1 & 0 & 0 & 1 & 0 & 1 & 1 & 0 & 1 & 3 & 2 & 2 & 3 & 2 & 2 & 2 & 2 & 3 & 2 & 3 & 2 & 2 & 3 & 3 & 2 \\[1.5ex]
            Transmitted symbol $s$                             & 0 & 0 & 0 & 0 & 0 & 1 & 0 & 1 & 0 & 0 & 0 & 0 & 0 & 1 & 0 & 1 & 0 & 0 & 1 & 0 & 0 & 0 & 1 & 1 & 0 & 0 & 0 & 1 & 0 & 1 & 1 & 1 \\
            \colrule
            R1: Received, 3 sym. offset                 &   &   &   & 0 & 0 & 0 & 0 & 0 & 1 & 0 & 1 & 0 & 0 & 0 & 0 & 0 & 1 & 0 & 1 & 0 & 0 & 1 & 0 & 0 & 0 & 1 & 1 & 0 & 0 & 0 & 1 & 0 \\[1.5ex]
            R2: Acceptance windows                            &   &   &   &   & 0 & 0 & 0 & 0 & 1 & 0 & 1 & 0 &   &   &   &   &   &   &   &   & 0 & 1 & 0 & 0 & 0 & 1 & 1 & 0 &   &   &   &   \\
            R3: Contrib. to $C$ for $\ell=1$             &   &   &   &   & 1 &-1 & 1 &-1 &-1 &-1 &-1 &-1 &   &   &   &   &   &   &   &   &   &   &   &   &   &   &   &   &   &   &   &   \\[1.5ex]
            R4: Shifted by $\delta_\mathrm{s}=1$              &   &   &   &   &   & 0 & 0 & 0 & 0 & 1 & 0 & 1 & 0 &   &   &   &   &   &   &   &   & 0 & 1 & 0 & 0 & 0 & 1 & 1 & 0 &   &   &   \\
            R5: Contrib. to $C$ for $\ell=2$             &   &   &   &   &   &   &   &   &   &   &   &   &   &   &   &   &   &   &   &   &   & 1 & 1 &-1 & 1 & 1 & 1 & 1 & 1 &   &   &   \\[1.5ex]
            R6: Shifted by $\delta_\mathrm{s}=1$              &   &   &   &   &   & 0 & 0 & 0 & 0 & 1 & 0 & 1 & 0 &   &   &   &   &   &   &   &   & 0 & 1 & 0 & 0 & 0 & 1 & 1 & 0 &   &   &   \\
            R7: Contrib. to $C$ for $\ell=3$             &   &   &   &   &   &   &   &   &   &   &   &   &   &   &   &   &   &   &   &   &   &-1 & 1 &-1 & 1 & 1 &-1 &-1 &-1 &   &   &   \\
        \end{tabular}
    \end{ruledtabular}
  \end{table*}

A typical QKD setup is depicted in Fig.~\ref{fig:setup}.
While the quantum channel requires a direct optical connection, our method allows for the classical data channel to be routed over a separate link with varying, unknown (but bounded) latencies up to hundreds of milliseconds.
In the following we assume that clock phase-locking is provided and focus on clock offset recovery.

Our method iQSync works as follows:
\begin{enumerate}
    \item Alice and Bob agree on the maximum level $\ell_\mathrm{max} \in \mathbb N$, which defines the maximum recoverable offset $\Delta_\mathrm{max}$, cf. Eq.~\eqref{eq:max-recoverable-offset}.
        They furthermore agree on the degree of interleaving $d_\mathrm{i}$.
        These values can be fixed and preconfigured.
    \item Alice sends a start signal to Bob via the data channel, providing Bob with a coarse marker of the transmission beginning.
        This is the only classical message needed for iQSync.
    \item At the same time, Alice starts the transmission of the synchronization pattern over the quantum channel, described in Section~\ref{sec:pattern-gen}.
    \item Upon reception of the start signal, Bob starts to acquire single photon detections and to analyze these according to the algorithm described in Section~\ref{sec:pattern-analysis}.
\end{enumerate}

Pattern generation and offset recovery are described in Sections~\ref{sec:pattern-gen} and \ref{sec:pattern-analysis} respectively.

An example Python implementation is given in the Supplemental Material \cite{supplementalMaterial}.

\begin{figure}
    \begin{algorithm}[H]
        \caption[justification=justified]{\label{alg:pattern-gen-iqsync}
        Pattern generation for iQSync.
        The pattern depends on the maximum level $\ell_\mathrm{max}$, and
        the degree of interleaving $d_\mathrm{i}$.
        Both values should be chosen such that
        the max. recoverable offset is sufficiently large, cf. eq.
        \eqref{eq:max-recoverable-offset}, and the detection statistics leads to a
        sufficiently high success probability, cf. Section~\ref{sec:success_probability}.
        An example pattern for $\ell_\mathrm{max} = 3$ and $d_\mathrm{i} = 2$
        is shown in Table~\ref{tab:symbols-iqsync-level-3-p-2}.
        LSB: Least significant bit; $\ll, \gg$: Bit shifts.
        }
        \hspace*{\algorithmicindent} \textbf{Input:} $\ell_\mathrm{max}$ \Comment{Max. level} \\
        \hspace*{\algorithmicindent} \textbf{Input:} $d_\mathrm{i}$ \Comment{Degree of interleaving}
        \begin{algorithmic}[1]
            \State $N_\ell \gets \ell_\mathrm{max} + 1$ \Comment{Number of levels}
            \State $N_\mathrm{g} \gets \lceil N_\ell / d_\mathrm{i} \rceil$ \Comment{Number of groups}
            \State $N_\mathrm{s,g} \gets 2^{\ell_\mathrm{max} + 1}$ \Comment{Number of symbols per group}
            \State $N_\mathrm{s} \gets N_\mathrm{s,g} \times N_\mathrm{g} $ \Comment{Total pattern duration (symbols)}
            \For{$k_\mathrm{s} \gets 0$ to $N_\mathrm{s} - 1$}
                \State $k_\mathrm{g} \gets k_\mathrm{s} \gg N_\ell$ \Comment{Index of current group}
                \State $\ell^- \gets k_\mathrm{g} \times d_\mathrm{i}$ \Comment{Min. level in group}
                \State $\ell^+ \gets \min(\ell^- + d_\mathrm{i} - 1, \ell_\mathrm{max})$ \Comment{Max. level in group}
                \State $\ell \gets $ randint $\in \{ \ell^-, ..., \ell^+ \}$ \Comment{Chosen level}
                \State $s \gets \mathrm{LSB}\pmb{(}(k_\mathrm{s} \ll 1) \gg \ell\pmb{)}$ \Comment{Select symbol}
                \State transmit symbol $s$
            \EndFor
        \end{algorithmic}
    \end{algorithm}
  \end{figure}
  \begin{figure}
    \begin{algorithm}[H]
        \caption{\label{alg:pattern-analysis-iqsync}
        Offset recovery for iQSync.
        The algorithm assumes binary PPM coding.
        As input it takes the maximum level $\ell_\mathrm{max}$, the degree of interleaving $d_\mathrm{i}$, and the timebin indices for which detections were obtained.
        It returns the clock offset $\delta$ in timebin precision.
        If Bob's clock is running ahead, $\delta > 0$, cf. Section~\ref{sec:pattern-analysis}.
        }
        \hspace*{\algorithmicindent} \textbf{Input:} $\ell_\mathrm{max}$ \Comment{Max. level} \\
        \hspace*{\algorithmicindent} \textbf{Input:} $d_\mathrm{i}$ \Comment{Degree of interleaving} \\
        \hspace*{\algorithmicindent} \textbf{Input:} $\mathcal{D}$ \Comment{Detection timebin indices} \\
        \begin{algorithmic}[1]
            \State $N_\ell \gets \ell_\mathrm{max} + 1$ \Comment{Number of levels}
            \State $N_\mathrm{s,g} \gets 2^{\ell_\mathrm{max} + 1}$ \Comment{Number of symbols per group}
            \State $k_\mathrm{s}^{-} \gets 2^{\ell_\mathrm{max} - 1}$
            \State $k_\mathrm{s}^{+} \gets N_\mathrm{s,g} - 2^{\ell_\mathrm{max} - 1}$
            \State $\delta \gets 0$ \Comment{Recovered offset}
            \State $k^{-} \gets 0$ \Comment{Index of first detection of current group}
            \For{$\ell \gets 0$ to $\ell_\mathrm{max}$} \Comment{Iterate over all levels}
                \State $C \gets 0$ \Comment{Counter}
                \State $g_\mathrm{req} \gets \lfloor \ell / d_\mathrm{i} \rfloor$
                \For{$k \gets k^{-}$ to $|\mathcal{D}| - 1$}
                    \State $k_\mathrm{s} \gets \lfloor \mathcal{D}(k) / 2 \rfloor$ \Comment{Current symbol index}
                    \State $g \gets k_\mathrm{s} \gg N_\ell$ \Comment{Current group}
                    \If{$g > g_\mathrm{req}$}
                        \State $g_\mathrm{req,next} \gets \lfloor (\ell + 1) / d_\mathrm{i} \rfloor$
                        \If{$g_\mathrm{req,next} > g_\mathrm{req}$}
                            \State $k^{-} \gets k$
                        \EndIf
                        \State \algorithmicbreak
                    \EndIf
                    \State $k_\mathrm{s,g} \gets$ the $N_\ell$ LSBs of $k_\mathrm{s}$ \Comment{Symb. idx. i. group}
                    \If{$k_\mathrm{s,g} < k_\mathrm{s}^{-}$ or $k_\mathrm{s,g} \ge k_\mathrm{s}^{+}$}
                        \State \algortihmiccontinue
                    \EndIf
                    \State $k_\mathrm{s, shifted} \gets [\mathcal{D}(k) + \delta] \gg 1$
                    \State $s_\mathrm{expected} \gets \mathrm{LSB}\pmb{(}(k_\mathrm{s, shifted} \ll 1) \gg \ell \pmb{)}$
                    \State $s \gets \mathrm{LSB}\pmb{(}\mathcal{D}(k) + \delta\pmb{)}$
                    \If{$s = s_\mathrm{expected}$}
                        \State $C \gets C + 1$
                    \Else
                        \State $C \gets C - 1$
                    \EndIf
                \EndFor
                \If{$C < 0$}
                    \State $\delta \gets \delta + 2^{\ell}$
                \EndIf
            \EndFor
            \If{$\delta > 2^{\ell_\mathrm{max}}$}
                \State $\delta \gets \delta - 2^{\ell_\mathrm{max} + 1}$
            \EndIf
            \State $\delta \gets - \delta$
            \State \Return $\delta$
        \end{algorithmic}
    \end{algorithm}
  \end{figure}

\subsection{\label{sec:pattern-gen}Pattern generation (Alice)}

The pattern transmitted by Alice can be generated during live operation with very few operations per symbol.

The synchronization pattern is divided into multiple \textit{levels}.
Each, except for level 0, is used to obtain one bit of the binary signed integer representation of the clock offset in symbol precision.
Level 0 consists of the symbol 0 only and is needed for systems which use more than one timebin to encode a symbol or qubit,
e.g. timebin-phase QKD systems
\cite{ruscaFinitekeyAnalysis1decoy2018,mollLinkTechnologyAlloptical2022},
to obtain proper symbol alignment.

Although different encodings can be used, symbols should not be encoded via the absence of optical pulses, e.g. via on-off keying, but via codings for which the channel attenuation does not introduce an asymmetry in the detected bit values, e.g. pulse-position modulation (PPM) or polarization coding.

Examples for patterns without and with interleaving are given in Table~\ref{tab:symbols-iqsync-level-2-p-1} and Table~\ref{tab:symbols-iqsync-level-3-p-2}, and general pattern generation, including level 0, is described in Algorithm~\ref{alg:pattern-gen-iqsync}.
Variables are initialized in lines 1-4, where $N_\ell$ is the total amount of levels, interleaved in $N_\mathrm{g}$ groups of $d_\mathrm{i}$ levels.
Each group has $N_\mathrm{s,g}$ symbols, leading to the total amount of $N_\mathrm{s}$ symbols.
A loop iterates over all symbols (line 5).
Out of the levels of the current group, a random level $\ell$ is chosen (lines 6-9), and the $\ell$'th least significant bit of the current symbol index is chosen for the transmitted symbol.

Recoverable offsets $\Delta$ in units of the symbol duration are limited by
\begin{align}
    - \Delta_\mathrm{max} \label{eq:max-recoverable-offset} \numberthis
    &\le
    \Delta < \Delta_\mathrm{max} - 1 \quad , \textrm{with} \\
    \Delta_\mathrm{max}
    &=
    2^{\ell_\mathrm{max} - 1} \, ,
\end{align}
cf. Fig.~\ref{fig:pattern_duration_over_max_offset}.
Hence, the total pattern duration scales as
$\mathcal{O}(n \log_2 n)|_{n=\Delta_\mathrm{max}}$ for $d_\mathrm{i}=1$
(no interleaving), and as
$\mathcal{O}(n)|_{n=\Delta_\mathrm{max}}$ for $d_\mathrm{i}=N_\ell$
(maximum interleaving).
A version similar to \cite{braithwaiteQuantumChannelSynchronization2020} is
obtained for $d_\mathrm{i} = 1$.

When choosing $d_\mathrm{i}$ as a power of two,
Algorithm~\ref{alg:pattern-gen-iqsync} only requires bit-shifts and additions
which enables straightforward implementation on low-level hardware,
like microcontrollers or FPGAs.

The time required for the transmission of the synchronization pattern is given by
\begin{align}
    T = N_\mathrm{s} \, t_\mathrm{s} \, ,
\end{align}
where $t_\mathrm{s}$ is the symbol duration.

\subsection{\label{sec:pattern-analysis}Offset recovery (Bob)}

To recover the clock offset, the detected pattern is evaluated at Bob in two steps.

First, Bob determines the shift necessary to align the detections to the
timebins, e.g. by employing a histogram.
This step defines the overall precision of the synchronization in the sub-timebin
regime.
Subsequently, this shift is applied to all detections.
Bob then converts all detection timestamps to timebin precision using remainder-free
division by the timebin duration.

Secondly, the clock offset in timebin granularity is reconstructed
using the dichotomic search Algorithm~\ref{alg:pattern-analysis-iqsync},
which assumes binary PPM coding with the bit one encoded as a late pulse.
The variable initialization in lines 1-2 is analog to Algorithm~\ref{alg:pattern-gen-iqsync}.
The total reconstructed offset in timebin granularity is stored in $\delta$.
The offset in symbol precision is given by $\delta_\mathrm{s} = \delta / 2$.

The outer loop (line 7-29) iterates over all levels, with the
currently analyzed level stored in $\ell$.
The counter variable $C$ stores the difference of the correct and wrong
symbols of the current level.
At the end of the evaluation of each level the counter is used to update
the offset $\delta$, if $C<0$ (lines 28-29).

The the inner loop (line 10-27) is used to determine $C$ by 
iterating over the detections belonging to the currently analyzed level.
This is achieved using the acceptance window defined by $k_\mathrm{s}^-$ and $k_\mathrm{s}^+$ in lines 3-4, which is applied to each group in lines 18-20.
The remaining detections are ensured to belong to the assumed level, given that the actual offset $|\Delta_\mathrm{actual}| < \Delta_\mathrm{max}$ symbols, cf. Eq.~\eqref{eq:max-recoverable-offset}.
The current symbol is shifted according to the already recoverd offset (line 21), and the counter $C$ is increased (decreased) in lines 26-29, in case the symbol corresponds to the expected/correct (unexpected/wrong) symbol.

Finally, lines 30-33 ensure that the reconstructed offset $\delta$ falls into the range $[-2\Delta_\mathrm{max}, 2\Delta_\mathrm{max}-1]$, and is positive in case Bob's clock is running ahead.

Note that this algorithm only uses integer variables and requires
no complex instructions like floating point operations, but only additions
and bit-shifts, since all multiplications and divisions by base-2 powers can be replaced by bit-shifts.
Thus, the algortihm is ideally suited for the implementation on FPGAs or microcontrollers.

A step-by-step example for offset recovery, using $\ell_\mathrm{max}=3$ and $d_\mathrm{i}=2$, is given in Table~\ref{tab:symbols-iqsync-level-3-p-2}, where it is assumed that Bob's clock runs ahead by 3 symbols ($\Delta = 3 t_\mathrm{s}$), cf. step R1 in Table~\ref{tab:symbols-iqsync-level-3-p-2}.
For simplicity, the example assumes that no signals are lost, and the recovery for $\ell=0$ is omitted, i.e. proper symbol alignment is assumed.
Application of the acceptance windows (lines 19-20 of Algorithm~\ref{alg:pattern-analysis-iqsync}) removes the detections from the first and last quarter of each group (R2).
Then, level 1 is evaluated (R3).
Therefore, the counter $C$ is increased (decreased) by 1 for each received symbol that matches (differs from) the expected symbol (lines 24-27 in Algorithm~\ref{alg:pattern-analysis-iqsync}).
Since $C=-4 < 0$ (sum over all values in R3), the recoverd offset $\delta$ is increased to $2^{\ell}=2$ (symbol offset $\delta_\mathrm{s}=\delta / 2 = 1$), cf. lines 28-29 in Algorithm~\ref{alg:pattern-analysis-iqsync}.
For the evaluation of level 2 the received symbols are then shifted by $\delta_\mathrm{s}$ (R4, line 23 in Algorithm~\ref{alg:pattern-analysis-iqsync}), and $C$ is again evaluated (R5).
Since counter $C=6 \ge 0$, the offset $\delta$ ($\delta_\mathrm{s}$) remains unchanged.
The evaluation of level 3 (R6, R7) leads to $C=-2 < 0$, and the offset is uptated to $\delta=10$ ($\delta_\mathrm{s}=5$).
Finally, since $\delta > 2^{\ell_\mathrm{max}} = 8$, the offset is reduced by $2^{\ell_\mathrm{max}+1}=16$, and its sign is inverted (lines 30-32 in Algorithm~\ref{alg:pattern-analysis-iqsync}), leading to $\delta=6$ ($\delta_\mathrm{s}=3$).
Hence, the correct offset is identified.
\section{\label{sec:success_probability}Success probability}

\begin{figure}
    \includegraphics[width=\linewidth]{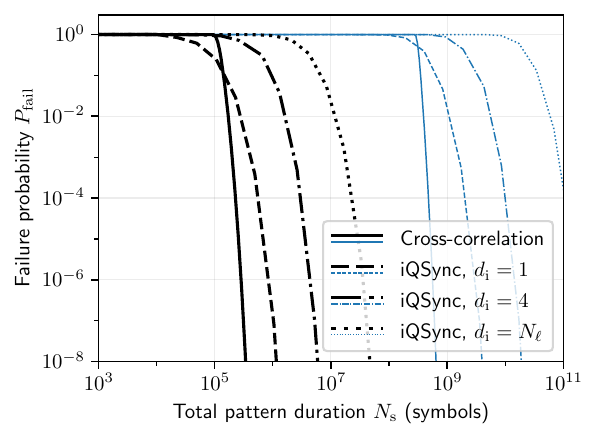}
    \caption{\label{fig:error_prob_over_pattern_duration}
    Calculated synchronization failure probability $P_\mathrm{fail}$ for two experimental
    conditions, depending on the total pattern duration $N_\mathrm{s}$.
    The thick black lines depict a scenario with high signal and low noise detection probability
    ($p_\mathrm{sig} = 10^{-3}$, $p_\mathrm{noise} = 10^{-7}$).
    The thin blue lines represent low signal and high noise detection probability
    ($p_\mathrm{sig} = p_\mathrm{noise} = 10^{-6}$).
    Cross-correlation refers to approaches utilizing a single FFT and the dual-step FFT approach as presented in \cite{calderaroFastSimpleQubitbased2020}.
    }
  \end{figure}

\begin{figure*}
    \includegraphics[width=\linewidth]{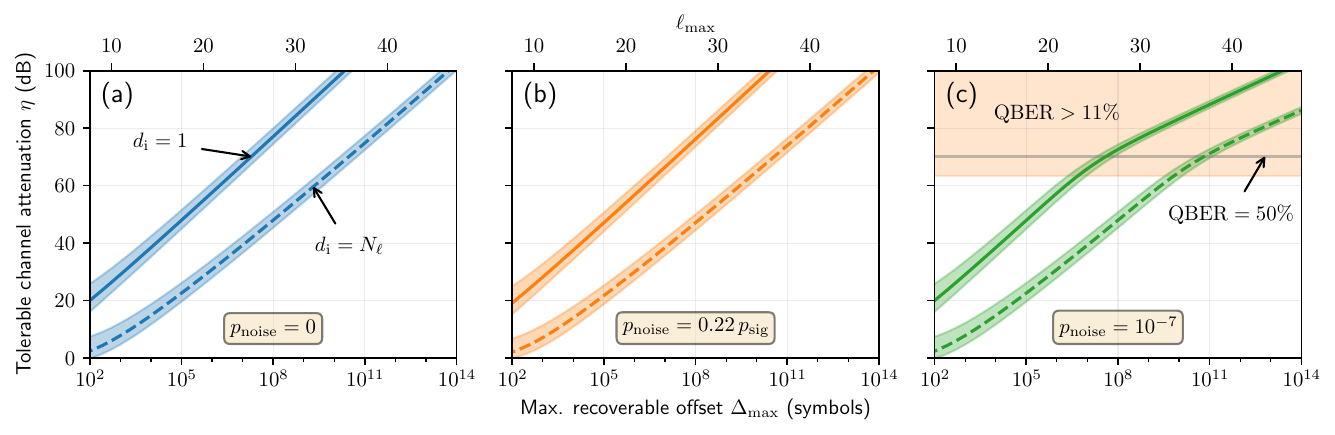}
    \caption{\label{fig:att_over_max_offset_3}
    Tolerable channel attenuation $\eta$ for iQSync, depending on the maximum recoverable offset $\Delta_\mathrm{max}$, the degree of interleaving $d_\mathrm{i}$, and the probability for a noise detection per symbol $p_\mathrm{noise}$.
    The solid (dashed) lines represent the settings with 50\% success probability for no (maximal) interleaving, with the lower (upper) edge of the shaded areas marking the 90'th (10'th) success percentile.
    The case of no noise is depicted in (a).
    The most pessimistic scenario for QKD is shown in (b), where for each configuration a QBER of 11\% is assumed.
    The case of $p_\mathrm{noise}=10^{-7}$ is shown in (c), representative for the detectors used during our experimental validation, cf. Section~\ref{sec:experiment}.
    The position of the kink depends on $p_\mathrm{noise}$ and always sits at the point, where the channel transmission equals $p_\mathrm{noise}$.
    The area with a $\mathrm{QBER} > 11\%$ is shaded red.
    }
\end{figure*}

\begin{figure}
    \includegraphics[width=\linewidth]{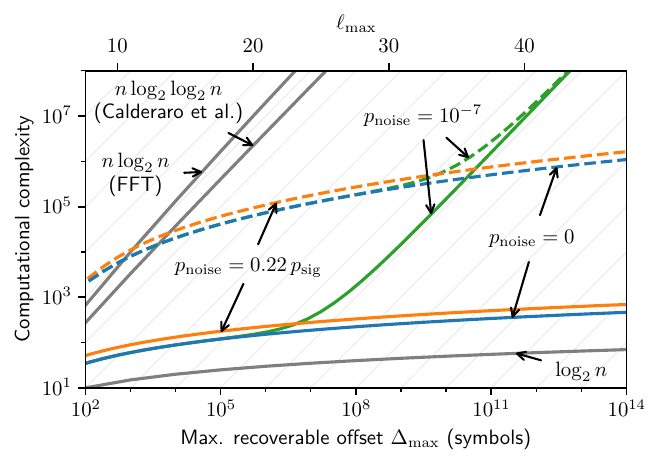}
    \caption{\label{fig:computational_complexity}
    TC of iQSync in terms of the number of iterations $\bar{N}_\mathrm{loop}$ over the inner loop of Algorithm~\ref{alg:pattern-analysis-iqsync} for different levels of noise, corresponding to the lines in Fig.~\ref{fig:att_over_max_offset_3}.
    The solid (dashed) lines represent the settings with 50\% success probability for no (maximal) interleaving.
    The three gray lines depict the TC of the FFT, the dual-step approach by Calderaro et al. \cite{calderaroFastSimpleQubitbased2020},
    and the theoretical limit $\log_2 n$, given by the number of bits of the binary representation of the offset.
    The light gray lines depict linear TC as a reference.
    }
\end{figure}

The success probability of the described method depends on the single-photon detection statistics.
To facilitate the straightforward identification of appropriate parameters $\ell_\mathrm{max}$ and
$d_\mathrm{i}$, an analytical expression for the success probability in terms of the signal (noise)
detection probability per symbol $p_\mathrm{sig}$ ($p_\mathrm{noise}$) is derived in
the following.
It is also the basis for the average-case TC analysis in Sec.~\ref{sec:computational_complexity}.

\begin{figure*}
    \includegraphics[width=\linewidth]{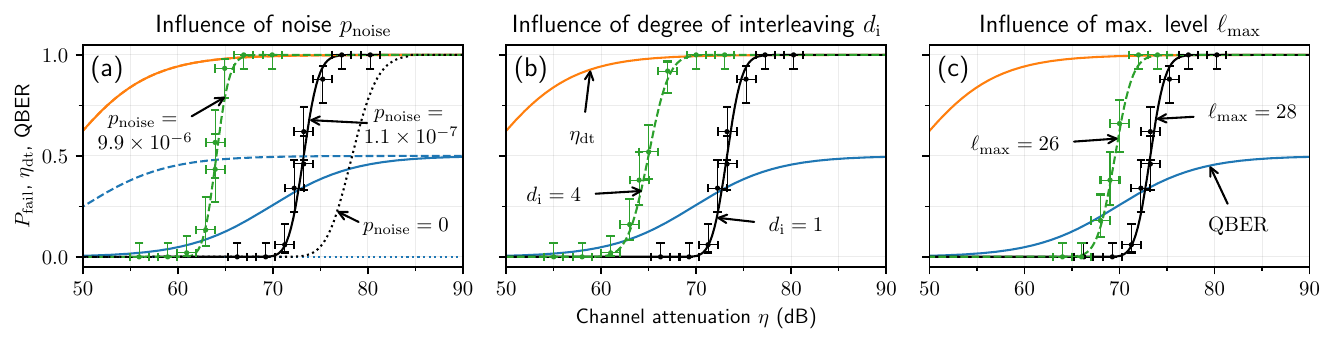}
    \caption{\label{fig:experimental_results}
    Comparison of experimental failure probabilities $P_\mathrm{fail}$
    and their modelling for different configurations.
    In (a) the solid black curve depicts $P_\mathrm{fail}$ for
    maximum level $\ell_\mathrm{max} = 28$, degree of interleaving $d_\mathrm{i} = 1$,
    and no added optical noise.
    The dashed green line shows $P_\mathrm{fail}$ for optical noise added on the
    detector.
    The dotted black curve shows the theoretical results for the case of no noise counts.
    The solid (dashed) blue curve depicts the expected QBER for the configuration
    without (with) added noise.
    The red curve shows the relative dead-time-induced reduction
    $\eta_\mathrm{dt}$ of the effective detector efficiency.
    The error bars represent the 95\%-CI for an uninformed prior.
    In (b) and (c) the dashed green line shows $P_\mathrm{fail}$ for an increased $d_\mathrm{i}$, and
    a reduced $\ell_\mathrm{max}$, respectively.
    }
\end{figure*}

The amount of symbols per group is given by
\begin{align}
    N_\mathrm{s,g}
    &=
    2^{\ell_\mathrm{max} + 1} \, .
\end{align}
The amount of signal detections, and hence the contribution to the counter
$C$ (cf. Algorithm~\ref{alg:pattern-analysis-iqsync}) per level is described by a Binomial distribution
\begin{align}
    C_\mathrm{sig}
    &\sim
    B\left(\frac{N_\mathrm{s,g}}{2}, p_\mathrm{sig} \frac{1}{d_\mathrm{i}} \right) \, ,
\end{align}
where the factor of $1/2$ was added, because the first and last quarter
of each group are removed by the acceptance window
(cf. Section~\ref{sec:pattern-analysis}).
The probability for a signal detection per symbol is denoted as $p_\mathrm{sig}$
and the factor $1/d_\mathrm{i}$ describes the fact that only signals from the currently analyzed level contribute constructively, while all other interleaved levels lead to random detections.
The influence of random detections onto the counter is given by
\begin{align}
    C_\mathrm{rand}
    &=
    N_\mathrm{rand,+} - N_\mathrm{rand,-} \, ,
\end{align}
with
\begin{align}
    N_{\mathrm{rand},\pm}
    &\sim
    B\left(\frac{N_\mathrm{s,g}}{2}, \frac{1}{2} p_\mathrm{rand} \right) \, ,
\end{align}
where
\begin{align*}
    \numberthis
    p_\mathrm{rand}
    &=
    1 - (1 - p_\mathrm{noise}) \big[ 1 - p_\mathrm{sig} (1 - 1/d_\mathrm{i}) \big] \\
    \overset{p_\mathrm{noise}, p_\mathrm{sig} \ll 1}&{\approx}
    p_\mathrm{noise} + p_\mathrm{sig} (1 - 1/d_\mathrm{i}) \, ,
\end{align*}
and $p_\mathrm{noise}$ denotes the probability for a noise detection per symbol.
Finally, the counter follows the distribution
\begin{align}
    C
    &\sim
    C_\mathrm{sig} + C_\mathrm{rand} \, .
\end{align}
Approximating the Binomial distributions by normal distributions via
\footnote{
The validity of this approximation should be checked for each context.
Requiring the $3\sigma$ interval of the distribution to be completely
contained within the interval $[0, 1]$ yields the requirement
$9(1-p) / (np) < 1$, which is satisfied for all measurements shown in
Fig.~\ref{fig:experimental_results}.
}
\begin{align}
    B(n, p)
    &\approx
    \mathcal N\pmb{(}np, np(1-p)\pmb{)}
    =
    \mathcal N(\mu, \sigma^2)
\end{align}
yields
\begin{align*}
    C
    &\sim
    \mathcal N(\mu_\mathrm{sig}, \sigma_\mathrm{sig}^2)
    + \mathcal N(\mu_\mathrm{rand,\pm}, \sigma_\mathrm{rand,\pm}^2) \\
    &\hspace{67pt} - \mathcal N(\mu_\mathrm{rand,\pm}, \sigma_\mathrm{rand,\pm}^2) \numberthis \\
    &=
    \mathcal N(\mu_\mathrm{tot}, \sigma_\mathrm{tot}^2) \, , \numberthis
\end{align*}
where
\begin{align}
    \mu_\mathrm{tot}
    &=
    \frac{N_\mathrm{s,g}}{2} \frac{p_\mathrm{sig}}{d_\mathrm{i}}
    =
    \mu_\mathrm{sig} \, , \\
    \sigma_\mathrm{tot}^2
    &= 
    \sigma_\mathrm{sig}^2 + 2 \sigma_{\mathrm{rand},\pm}^2 \, , \\
    \sigma_\mathrm{sig}^2
    &=
    \frac{N_\mathrm{s,g}}{2} \frac{p_\mathrm{sig}}{d_\mathrm{i}} \Big( 1 - \frac{p_\mathrm{sig}}{d_\mathrm{i}} \Big) \, , \\
    \sigma_{\mathrm{rand},\pm}^2
    &=
    \frac{N_\mathrm{s,g}}{2} \frac{p_\mathrm{rand}}{2} \Big( 1 - \frac{p_\mathrm{rand}}{2} \Big) \, .
\end{align}

The probability for successful analysis of one level is then found as
\begin{align}
    P_{\mathrm{success},1}
    =
    P(C > 0)
    \approx \Phi \Big( \frac{\mu_\mathrm{tot}}{\sigma_\mathrm{tot}} \Big) \, ,
\end{align}
where $\Phi$ is the cumulative distribution function of the standard normal
distribution.
This leads to the probability
\begin{align}
    \label{eq:p_success}
    P_\mathrm{success}
    =
    P_{\mathrm{success},1}^{N_\ell}
\end{align}
for the whole synchronization to succeed.

For a given link, an increase of the degree of interleaving $d_\mathrm{i}$ leads to shorter synchronization patterns at the expense of higher failure probabilities, cf. Fig.~\ref{fig:pattern_duration_over_max_offset}, Fig.~\ref{fig:error_prob_over_pattern_duration}, and Fig.~\ref{fig:experimental_results}(b).
\section{\label{sec:computational_complexity}Time complexity}

\begin{figure*}
    \includegraphics[width=\linewidth]{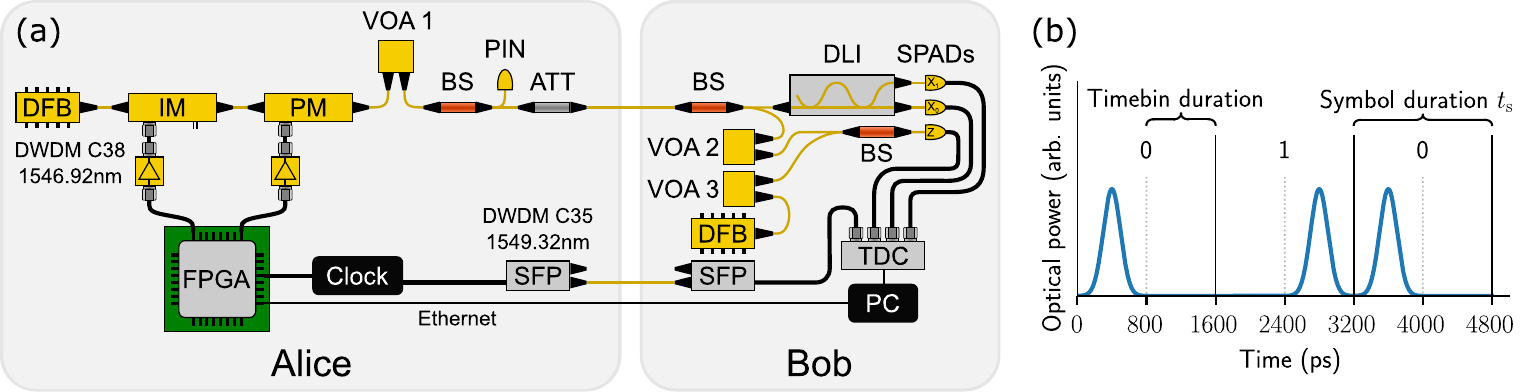}
    \caption{\label{fig:experimental_setup}
    Experimental setup.
    In (a) the systems for Alice and Bob are shown, as they were used for verifying the analytically derived success
    probabilities (Section~\ref{sec:success_probability}).
    The clock beat was transmitted via fiber.
    Using VOA 2, the channel attenuation was applied in the $\mathsf{Z}$-basis only,
    such that the $\mathsf{X}$-basis QBER could be used to independently check
    for synchronization success.
    Additional optical noise was added directly in front of the $\mathsf{Z}$-basis
    detector via a second laser and could be tuned via VOA 3.
    The symbol encoding, using binary PPM, is shown in (b) for the example symbol sequence "010".
    For the experiments a symbol duration $t_\mathrm{s} = 1600\,\mathrm{ps}$ was used.
    FPGA: field-programmable gate array;
    SFP: Small form-factor pluggable;
    DFB: Distributed-feedback laser;
    IM: Intensity modulator;
    PM: Phase modulator;
    VOA: Variable optical attenuator;
    BS: Beam splitter;
    PIN: Photodiode;
    ATT: Fix attenuator;
    DLI: Delay-line interferometer;
    SPAD: Single-photon avalanche detector;
    TDC: Time digital converter.
    }
\end{figure*}

The TC for pattern generation is given by the pattern duration, i.e. $\mathcal{O}(n \log_2 n)$ for $d_\mathrm{i}=1$, and $\mathcal{O}(n)$ for $d_\mathrm{i}=N_\ell$, cf. Eq.~\eqref{eq:max-recoverable-offset} and Algorithm~\ref{alg:pattern-gen-iqsync}.

For the recovery algorithm the TC can be analyzed numerically, using the previously derived analtyical expression for the success probability, leading to the channel attenuation
\begin{align}
    \eta = \eta(\ell_\mathrm{max}, d_\mathrm{i}, p_\mathrm{noise}, P_\mathrm{success}) \, .
\end{align}
Application of the constraint $P_\mathrm{success}=0.5$, used to establish a fair comparison, produces the lines shown for different configurations in Fig.~\ref{fig:att_over_max_offset_3}.
Due to the inherent statistical nature of the presented offset recovery, we quantify the average-case TC.
Therefor we employ the average number of iterations over the inner loop in Algorithm~\ref{alg:pattern-analysis-iqsync}, i.e.
\begin{align}
    \bar{N}_\mathrm{loop}
    &=
    p_\mathrm{det} N_\mathrm{s,g} N_\ell \quad \text{, with} \\
    p_\mathrm{det}
    &=
    1 - (1-p_\mathrm{sig})(1-p_\mathrm{noise}) \, .
\end{align}
Evaluating $\bar{N}_\mathrm{loop}$ along the lines of Fig.~\ref{fig:att_over_max_offset_3} results in the TCs depicted in Fig.~\ref{fig:computational_complexity}.

Sublinear TC is achieved for $p_\mathrm{noise} = 0$, as well as for $p_\mathrm{noise} = 0.22 \, p_\mathrm{sig}$, which corresponds to a QBER of $11\,\%$, the upper limit for QKD.
For $p_\mathrm{noise}=0$ the TC can be fitted by the poly-logarithmic function
\begin{align}
f(n)
=
a (\log_2 n)^b \, .
\end{align}
With $a \approx 2.9, b \approx 1.3$ ($a \approx 4.9, b \approx 3.2$) for no (max.) interleaving the relative deviation is bounded by $2.9\,\%$ ($5.1\,\%$) over the range shown in Fig.~\ref{fig:computational_complexity}, which entails all offsets relevant to practical QKD implementations.

Only for constant values of $p_\mathrm{noise} > 0$ the TC takes super-linear complexity in the regime past the point with a QBER of $11\,\%$, given by the intersection with the curve for $p_\mathrm{noise} = 0.22 \, p_\mathrm{sig}$.
However, this regime is of little importance for QKD, since it does not allow for the generation of secure keys.

Hence, the clock offset recovery Alogirthm~\ref{alg:pattern-analysis-iqsync} has poly-logarithmic average-case TC for the parameter regime relevant for QKD.
\section{\label{sec:experiment}Experiment}

We experimentally validated our method and compared it with the analytically
derived success probabilities, cf. Section~\ref{sec:success_probability}.
Therefor, iQSync was implemented in our
timebin-phase QKD system, employing the finite-size 1-decoy timebin-phase BB84 protocol \cite{mollLinkTechnologyAlloptical2022,ruscaFinitekeyAnalysis1decoy2018},
using two timebins per qubit and a qubit transmission rate of $625\,\mathrm{MHz}$.

The system was used in the experimental configuration as shown in Fig.~\ref{fig:experimental_setup}.
At Alice, the pattern was generated in real-time on an FPGA (Xilinx Ultrascale+ VU13P),
using a synthesized AES-CTR module for the random level choice,
if $d_\mathrm{i} > 1$, cf. Algorithm~\ref{alg:pattern-gen-iqsync}.
The transmitted mean photon number per symbol was set to 1 during the synchronization.

The channel attenuation was realized using a variable optical attenuator (VOA)
in front of the $\mathsf{Z}$-basis detector, which was used for the synchronization.
This setup allowed to use the $\mathsf{X}$-basis QBER to reliably identify successful synchronizations,
even at very high channel attenuations for which QBERs near 50\%
were observed in the $\mathsf{Z}$-basis.
Stirling-cooled SPADs (ID Quantique ID230) with a dead-time
$\tau_\mathrm{dt}=96.3 \, \mathrm{\mu s}$ and single photon detection efficiency
$\eta_\mathrm{det}=0.154$ were used.
Offset recovery (Algorithm~\ref{alg:pattern-analysis-iqsync}) was implemented in a few lines of \texttt{C++} code and evaluated during real-time operation.

Comparative outcomes between experimental failure probabilities, $P_\mathrm{fail}$, and their modeling are depicted in Fig.~\ref{fig:experimental_results}.
The baseline condition is represented by the solid black curves, established through a maximum level of $\ell_\mathrm{max} = 28$, a degree of interleaving of $d_\mathrm{i} = 1$, and the absence of additional optical noise, leading to $p_\mathrm{noise} = 1.1 \times 10^{-7}$, caused by detector dark counts.
The resulting pattern has a duration of $24.9 \, \mathrm{s}$ and allows to recover offsets of up to $\Delta_\mathrm{max} = 215 \, \mathrm{ms}$.

Fig.\ref{fig:experimental_results}(a) demonstrates the introduction of optical noise to the detector, leading to a noise detection probability of $p_\mathrm{noise} = 9.9 \times 10^{-6}$.
Fig.\ref{fig:experimental_results}(b) denotes a condition wherein the degree of interleaving was increased to $d_\mathrm{i} = 4$ and Fig.~\ref{fig:experimental_results}(c) depicts a situation where the maximum level was decreased to $\ell_\mathrm{max} = 26$.
For each observation point, the system was started 50 times, and successful synchronizations were counted.
The channel attenuation, as shown on the horizontal axis, includes all losses between Alice and the $\mathsf{Z}$-basis detector, and encompasses the detector efficiency as well as its efficiency reduction induced by dead-time.

For $\ell_\mathrm{max}=28$ and $d_\mathrm{i}=4$ the synchronization pattern spanned $6.9\,\mathrm{s}$, leading to $4372 \pm 63$ detections with reliable recovery (49 out of 50 runs) of offsets up to $215\,\mathrm{ms}$ at $61.0\,\mathrm{dB}$ channel attenuation, cf. Fig.~\ref{fig:experimental_results}(b).

For $\ell_\mathrm{max}=28$ and $d_\mathrm{i}=1$ the synchronization pattern spanned $24.9\,\mathrm{s}$, leading to $3056 \pm 62$ detections with reliable recovery (47 out of 50 runs) of offsets up to $215\,\mathrm{ms}$ at $71.2\,\mathrm{dB}$ channel attenuation, cf. Fig.~\ref{fig:experimental_results}(b).

Furthermore, iQSync has successfully been used by our QKD system in combination
with fiber and free-space links (publications in preparation).

In summary, these experimental outcomes demonstrate the method's performance and robust conformity with the theoretical modeling.
As a result, the theoretical model outlined in Section~\ref{sec:success_probability} can be effectively utilized to discern optimal values for $\ell_\mathrm{max}$ and $d_\mathrm{i}$ for a given link configuration, and our synchronization method can be used for channel attenuations exceeding $70\,\mathrm{dB}$.
\section{\label{sec:conclusion}Conclusion}

We introduced iQSync, a synchronization method suited for low-level hardware implementations.
iQSync works by successively identifying the binary representation bits of the clock offset, starting from the least significant bit, and interleaving the requisite patterns to reduce the total pattern duration.
The pattern is determined by two parameters: its maximum level $\ell_\mathrm{max}$, defining the maximum recoverable offset, and the degree of interleaving $d_\mathrm{i}$, by which the total pattern duration can be reduced.
For the random level interleaving we used AES-CTR random number expansion, but simpler algorithms should suffice.

We derived an analtytical expression for the success probability, and the average-case time complexity of the recovery algorithm was found to be poly-logarithmic for all settings relevant to QKD using the BB84 protocol, i.e. QBERs below $11\,\%$.
Furthermore, we demonstrated excellent agreement between the analytical results and experimental measurements for different link and parameter configurations.
Reliable synchronization, requiring only a few thousand iterations over a simple loop, was demonstrated for total channel attenuations exceeding $70\,\mathrm{dB}$.

Greater channel losses could readily be compensated by using more than one photon per symbol during the synchronization, enabling operation even under most adverse conditions
\cite{boaronSecureQuantumKey2018}.
In prepare-and-measure QKD systems the number of photons per qubit can often readily be adjusted using Alice's VOA.
Alternatively, the tolerable channel attenuation for a given $\ell_\mathrm{max}$ can also be increased by prolongation of the level patterns, which requires a slight modification of Algorithms~\ref{alg:pattern-gen-iqsync} and \ref{alg:pattern-analysis-iqsync}.
Hence, our method is well suited for prepare-and-measure QKD systems, even for noisy channels with high attenuation.

Furthemore, iQSync could also be of interest for other applications, such as LIDAR, optical deep space communication, or communication out of line of sight.

The presented experimental results and successful operation during the last three years demonstrate the reliability and applicability of iQSync to real communication networks.
\section{\label{sec:acknowledgements}Acknowledgements}

We thank Andy Schreier for valuable feedback on this manuscript.

Large Language Models were used during writing to polish and lightly edit some passages.

This research was conducted within the scope of the project QuNET,
funded by the German Federal Ministry of Education and Research (BMBF)
in the context of the federal government's research framework in
IT-security “Digital. Secure. Sovereign.”.
\FloatBarrier
\section{\label{sec:author_contributions}Author Contributions}

N.W. proposed and directed the research.
J.K. developed the synchronization method with guidance from N.W.
J.K. conceived and implemented the algorithms, including the example Python implementation in the Supplemental Material \cite{supplementalMaterial}.
J.H. implemented the transmission algorithm on the FPGA.
J.K. performed the experiments, evaluated the measurement data, analytically derived the success probability, and conducted the time complexity analysis.
R.F. acquired the funding.
J.K. wrote the manuscript.
All authors discussed the results and reviewed the manuscript.

\bibliography{literature}

\end{document}